# A Novel Stability Improvement Method of S-Band Magnetron Systems Based on Its Anode Current Feature

Shaoyue Wang, *Graduated Student Member, IEEE*, Yan Zhao, Xiaojie Chen, and Changjun Liu, *Senior Member, IEEE*

*Abstract*—Magnetrons are widely used as a high-efficiency, large-power, and low-cost microwave source. This paper proposes a method to improve the magnetron output characteristics based on regulating its anode current by a novel active distortion eliminator. It takes advantage of the constant current state of MOSFETs to suppress the anode current ripple and eliminate the magnetron output ripples. An improved switch-mode power supply is applied to a 1-kW commercial S-band magnetron in the experiment. The peak-to-peak anode current ripple is reduced from 31 mA to 7 mA. The spectral width of the free-running magnetron is decreased from 246 kHz to 87 kHz. In the injection locking state, the spurious noise is significantly reduced with a maximum suppression strength of 23 dB at a 100 Hz offset. The peak-to-peak value of the magnetron's magnitude and phase jitter are suppressed to 15% and 16% of their original values at a power injection ratio of 0.13. This method provides a novel approach for stabilizing the output phase of an injection-locked magnetron, making it promising for applications in phased arrays, wireless power transfer, particle accelerators, large-power communication, and other scenarios that require high phase stability.

## I. INTRODUCTION

Microwave applications such as particle accelerators, phase-controlled arrays, wireless power transmission, and simultaneous wireless information and power transfer always require large-power microwave sources with excellent performance [1]-[5]. Magnetrons are widely used microwave sources and have attracted much attention from researchers because of their advantages of low cost, low weight, high direct current (dc) to radio frequency (RF) conversion efficiency, and high output power. However, the magnetrons' poor phase stability and randomized output spectral band lead to challenges in beam controlling, power combining, and minimized interference to adjacent frequency bands, and so on, which are required in the abovementioned areas [6]. These defects restrict the magnetron applications greatly.

Researchers have conducted extensive studies to overcome the drawbacks of magnetrons [7]-[9]. The injection-locking technique is an efficient way to improve the output microwave quality of a magnetron [10]. It allows complete control over a magnetron's output microwave frequency and phase by injecting an external high-precision and low-power signal. However, the performance of an injection-locked magnetron is also closely related to the dc power supply and its intrinsic characteristics.

Optimizing the field distribution characteristic of a magnetron's resonant cavity is a method to improve its output quality. Neculaes *et al.* successfully restored the axial symmetry of the magnetic field and reduced the noise near the carrier frequency by perturbing magnets [11]-[13]. Another way is to improve the performance of magnetrons' dc power supplies. Chen *et al.* added a passive filter module to suppress the power supply's voltage ripples and significantly enhance the output microwave purity of a 20-kW magnetron [14]. T.-S. Lee *et al.* proposed a dc power supply with gain-frequency tracking control. The heating effects and cavity power distribution were well regulated and achieved excellent constant power operation [15].

The phase-locked loop (PLL) technique is also an efficient way to control the output phase of a magnetron precisely. Yang. *et al.* successfully controlled a 5.8GHz magnetron's output phase by a PLL and stabilized its output power by adjusting the anode current. The magnetron has a 1% power stability and a nearly ±1° phase accuracy [16]. The S. T. Han *et al.* combined a low-ripple, high-voltage dc power supply with a phase-locked circuit and successfully obtained phase stability at 0.3° peak-to-peak [17]. Chen *et al.* proposed a magnetron power combining the system with a solid-source-like performance by injection-locking and close-loop compensation, achieving phase jitter as low as ±0.5° [18].

Improving the microwave output system setup is another aspect. Liu *et al.* studied the load-pull effect of magnetrons and significantly enhanced the magnetron's injection ratio and locking bandwidth by tuning the magnetron's output load [19]. Chen *et al.* investigated the performance of an injection-locked magnetron with various load reflection levels. They found that a proper-mismatched system would suppress magnetrons' sideband energy, thereby reducing phase noise [20]. However, achieving precise phase output only through the power supply makes the system complex and bunk. Capacitor or inductor filter modules are always passive and low-complexity. However, they are expensive and large-volume and cannot eliminate all the spurious noise caused by the power supply. The PLL circuit is active and high-

This work was supported in part by the National Natural Science Foundation of China (NSFC) under Grant U22A2015 and 62071316.

S. Wang, Y. Zhao, and C. Liu are with the School of Electronics and Information Engineering, Sichuan University, Chengdu 610064, China (*Corresponding author: Changjun Liu*. e-mail: cjliu@ieee.org).

X. Chen is with the High People's Court of Guangxi Autonomous Region, Nanning 530028, China.

performance, and it will increase the system's complexity and require a sophisticated design process. Besides, C.-S. Ha *et al.* found a magnetron system with PLL still suffered the influence of the power supply's ripple [21].

As a closely related factor because of the frequency pushing effect, anode current is essential for improving magnetrons' output quality [22]. Mitani *et al.* found the filament current of a DC power supply causing magnetrons' anode current fluctuation, then they successfully reduced the magnetron's output noise and realized a narrowband spectrum by cutting off the filament current [23][24]. Imran Tahir *et al.* controlled the current of a switch-mode power supply to drive magnetrons as current-controlled oscillators in a phase-locked loop [25]. These studies indicate that approaching power supply improvements from the perspective of current characteristics is an effective strategy. But in recent years, there have been relatively few methods for improving the anode current of magnetrons.

In this paper, we proposed a novel anode current ripple elimination technique that utilizes a current control method to improve magnetrons' output quality. The technique is realized by an active distortion eliminator (ADE) with compact size and high performance. The ADE can suppress the ripples caused by the anode current's power-line or switching noise and eliminate the magnetron's microwave output distortion. A theoretical analysis of the output spectrum of magnetrons under the influence of current is carried out. A numerical study of magnetrons' output influenced by the current ripples is also conducted.

A switch-mode power supply is improved, and a series of verification experiments are conducted to verify the proposed method. An S-band 1-kW commercial magnetron is measured. Both the free-running and injection-locked magnetrons' output noise levels are suppressed significantly. The magnitude and phase jitter of the injection-locked magnetron is as low as 0.07dB and 0.32° (peak-to-peak value) without applying any closed-loop phase control circuit on the magnetron's output microwave. This method is simple and easy to implement with significant effects. The advantages are compact size, low cost, and simple structure. This low-cost solution is suitable for many applications where improving the magnetron output spectrum is desired.

## II. THEORETICAL AND NUMERICAL ANALYSIS

### A. Free-Running Magnetrons

A magnetron system always includes a power supply system, resonant cavity, output antenna, and load. According to the pioneer research by G. B. Collins and J. C. Slacter [26][27], a magnetron operating in a single mode can be equivalent to a parallel RLC resonant network, and the load, resonant cavity and dc power supply influence its electron stream characteristics. We use the equivalent circuit in Fig. 1(a) to represent the magnetron. The *R*, *L*, and *C* in the diagram represent the equivalent resistance, inductance, and capacitance of the magnetron resonant cavity, respectively. They are related to the magnetron structure. *G*+j*B* represents the magnetron equivalent load admittance. $V_{dc}$ and $I_{dc}$ refer to the magnetron anode voltage and current supplied by the dc power source during the operation. $V_{RF0}$ and $I_{RF0}$ are the magnetron's output signal's RF voltage and current amplitudes. The term $g+jb$ is the equivalent admittance of the electrons between the magnetron's cathode and anode, which is influenced by various factors.

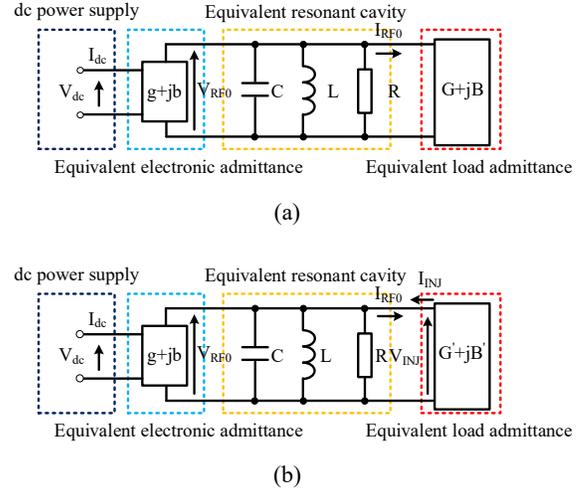

Fig. 1. The equivalent circuit of a magnetron. (a) a free-running magnetron. (b) an injection-locked magnetron.

We assumed that when the magnetic field of the magnetron is constant, they can be expressed as functions of the dc voltage $V_{dc}$ and $I_{dc}$. When the magnetron operates stably, the equivalent circuit satisfies

$$g + jb = \frac{1}{R} + j\omega_c C + \frac{1}{j\omega_c L} + G + jB \quad (1)$$

where $\omega_c$ is the output frequency of the magnetron. By rearranging (1), we have

$$g = \frac{1}{R} + G \quad (2)$$

$$b = \omega_c C - \frac{1}{\omega_c L} + B, \quad \omega_0^2 = \frac{1}{LC} \quad (3)$$

where $\omega_0$ is the native resonant frequency of the magnetron's resonant cavity, it is solely related to the structure of the magnetron itself. By solving (3), the free-running frequency of the magnetron is obtained [28]:

$$\omega_c = \frac{b-B}{2C} + \omega_0 \sqrt{\frac{(b-B)^2}{4\omega_0 C^2} + 1} \quad (4)$$

According to J. C. Slater's work, the equivalent electronic admittance of the magnetron is given as follows:

$$g = \frac{1}{R}\left(\frac{V_{dc}}{V_{RF0}} - 1\right) \quad (5)$$

$$b = b_0 + g \tan\alpha \quad (6)$$

The load admittance for a free-running magnetron is

$$G + jB = \frac{I_{RF0} e^{j\omega_c t}}{V_{RF0} e^{j\omega_c t}} \quad (7)$$

For a magnetron, there exists a mutual interaction between its $V_{dc}$ and $I_{dc}$. When one of them is changed, the other one also varies accordingly. The anode voltage affects the motion of electrons emitted from the magnetron's cathode. It will determine the shape and phase of the electron spokes and, consequently, the number of electrons reaching the anode, leading to control of its anode current. Conversely, when the anode current changes, it impacts the phase of the electron spokes and its internal equivalent capacitance. At this point, as the magnetron serves as a load for the power supply, its impedance characteristic changes, thereby causing variations in the anode voltage. Therefore, we express the magnetron anode voltage as

$$V_{dc} = f(I_{dc}) \tag{8}$$

Due to the relationship between load admittance and anode current, $g$ and $b$ are rewritten as $g(I_{dc})$ and $b(I_{dc})$, respectively. We have

$$I_{dc} = I_0 + \Delta I \tag{9}$$

where $I_0$ is the ideal anode current without any ripples, and $\Delta I$ is the spurious component of the anode current, which is a function of time. Then, combining (4) and (9), expanding the result of $I_0$, we have

$$\omega_c = \omega' + \Delta \omega \approx \omega_0 + \frac{b(I_0) - B}{2C} + \frac{\Delta I}{2C} \frac{\partial b(I_{dc})}{\partial I_0} \tag{10}$$

where $\omega'$ is the magnetron output frequency without any interference. Approximately, by taking the mean square value of $\Delta \omega$, we can obtain the variation of a free-running magnetron's output frequency

$$\overline{(\Delta \omega)}^2 = \overline{\Delta I}^2 \left( \frac{1}{2C} \frac{\partial b(I_{dc})}{\partial I_0} \right)^2 \tag{11}$$

Similarly, combining (2), (5) and (8), according to the circuit theory, we have $Q_{ext} = \omega_0 C/G$, $Q_0 = \omega_0 CR$, where $Q_{ext}$ and $Q_0$ are the magnetron's external and intrinsic quality factors, respectively. We obtain the output voltage of a free-running magnetron

$$V_{RF0} = \frac{f(I_{dc})}{2RC\gamma}, \quad \gamma = \omega_0 \left( \frac{1}{Q_0} + \frac{1}{2Q_{ext}} \right) \tag{12}$$

where $\gamma$ is defined as the growth parameter of the magnetron [22]. As the ripple of the anode current increases, the mean square value of $\Delta \omega$ also increases, and the stability of the magnetron's output is worse. Thus, for a free-running magnetron, suppressing its anode current fluctuations will enhance the stability of the magnetron's output frequency and reduce the spurious noise.

*B. Injection-Locked Magnetrons*

The equivalent circuit of an injection-locked magnetron is presented in Fig. 1(b). When an injection locking signal is applied to a magnetron, the magnetron equivalent load admittance varies due to the signal injection. The voltage and current of the injection source are $V_{inj}$ and $I_{inj}$, respectively. The equivalent load admittance of the magnetron is

$$G' + jB' = \frac{I_{RF0} e^{j\omega_c t} - I_{inj} e^{j\omega_{inj} t}}{V_{RF0} e^{j\omega_c t} + V_{inj} e^{j\omega_{inj} t}} = \frac{I_{RF0}}{V_{RF0}} \left( \frac{1 + \frac{I_{inj}}{I_{inj}} e^{j(\omega_{inj} - \omega_c)t}}{1 + \frac{V_{inj}}{V_{RF0}} e^{j(\omega_{inj} - \omega_c)t}} \right) \tag{13}$$

$$\approx G \left[ 1 - 2\rho e^{j\theta} \right] = G - 2G\rho \cos\theta - j2G\sin\theta$$

where $\theta = (\omega_{inj} - \omega_c)t$ is the phase difference between the magnetron's output and injected signal, and the injection ratio $\rho$ is defined as $\rho = \sqrt{P_{inj}/P_{RF0}} = V_{inj}/V_{RF0}$. Then, by substituting (13) into (1), we obtain

$$g + jb = \frac{1}{R} + j\omega_c C + \frac{1}{j\omega_c L} + 2G(1 - \rho\cos\theta - j\rho\sin\theta) \tag{14}$$

Similar to the derivation for free-running magnetrons, the imaginary part of (14) is

$$b = \omega_c C - \frac{1}{\omega_c L} - 2G\rho\sin\theta, \quad \omega_0^2 = \frac{1}{LC} \tag{15}$$

Here, we have the assumption of $\omega_{inj} \approx \omega_c$. Then, similarly, by solving (15) and omitting higher-order infinitesimal terms, we obtain

$$\omega_c' \approx \omega_c(\Delta I) - \frac{\omega_0 \rho}{Q_{ext}} \sin\theta \tag{16}$$

where $\omega_c'$ is the actual output frequency of the magnetron under the injected signal. It is time-varying. When injection locking occurs, $\omega_c'$ eventually becomes equal to the frequency of the injected signal. $\omega_c(\Delta I)$ is different from the free-running situation due to the change of load admittance caused by injection locking:

$$\omega_c(\Delta I) \approx \omega_0 + \frac{b(I_0)}{2C} + \frac{\Delta I}{2C} \frac{\partial b(I_{dc})}{\partial I_0} \tag{17}$$

Then, we subtract $\omega_{inj}$ from both sides of (16) and set $d\theta/dt = \omega_{inj} - \omega_c'$, which can be facilitated as:

$$\frac{d\theta}{dt} = \omega_{inj} - \omega_c(\Delta I) - \frac{\omega_0}{Q_{ext}} \rho\sin\theta \tag{18}$$

Here, we disregard the control effects of injection locking on the output voltage amplitude of the magnetron, which is expressed as:

$$\frac{d\theta}{dt} = \omega_{inj} - \omega_c(\Delta I) - \frac{2RC\gamma V_{inj}}{f(I_{dc})Q_{ext}} \sin\theta \tag{19}$$

It shows the state of the injection-locked magnetron. Although the frequency and phase of the magnetron output are synchronized with the external signal, ripples in the magnetron's anode current will still influence the final output phase.

Enhancing the characteristics of the magnetron's anode current will effectively improve the phase characteristics of the injection-locked magnetron, and it plays a crucial role in improving the performance of both free-running and injection-locked magnetrons.

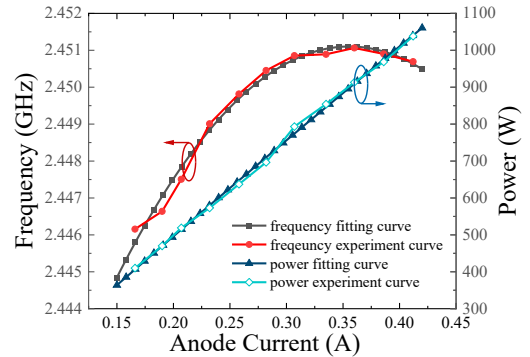

Fig. 2. Experiment and polynomial fitting curves of magnetron's power and frequency versus anode current.

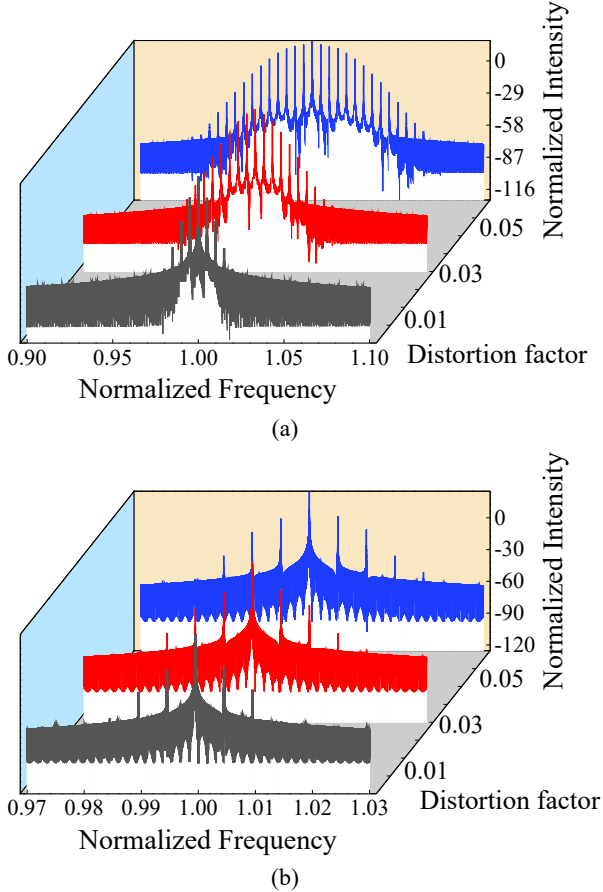

Fig. 3. Results of numerical analysis of magnetron spectrum under different anode current ripple factors. (a) Free-running states. (b) Injection-locked states.

### C. Numerical Analysis of Frequency and Phase Characteristics of Magnetrons

We conducted a series of numerical calculations to assist the theoretical analysis and make the results more intuitive. In the previous section, we derived the relationship between the magnetron's frequency, power, and output phase with its anode current. However, in these relationships, the magnetron's susceptance $b$, resonant cavity capacitance $C$, resistance $R$, and the derivative of the susceptance $b$ at $I_0$ are challenging to obtain through simulations and calculations. Therefore, in this section, we directly obtained the values of magnetron output power $P_{RF0}$ and frequency $\omega_c$ at various anode currents $I_{dc}$ through experiments.

In the experiments, a Panasonic 2M210-M1 1-kW S-band magnetron is driven by a controllable low-ripple DC power supply. Its output frequency and power were monitored using a spectrum analyzer and a power meter. Fig. 2 shows the curves of measured and fitted magnetron power and frequency as a function of anode current. In this case, we obtained the relationship between frequency and current through polynomial fitting as follows:

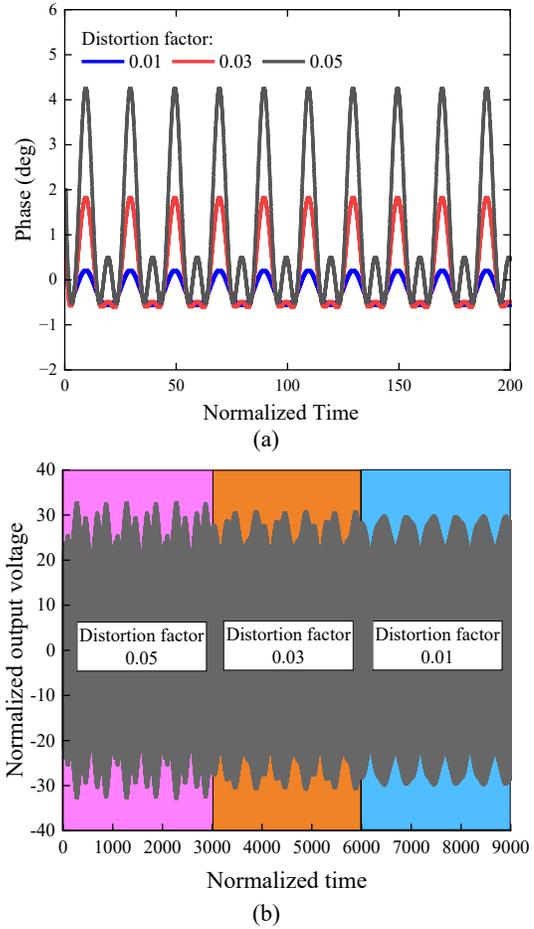

Fig. 4. Numerical analysis results of (a) output phase and (b) output voltage of injection-locked magnetrons under different anode current ripple factors.

$$f_c(I_{dc}) = a_0 + a_1 I_{dc} + a_2 I_{dc}^2 \quad (20)$$

where $a_0$=2.4324, $a_1$=0.1053, $a_2$= −0.1484, and:

$$P_{RF0}(I_{dc}) = b_0 + b_1 I_{dc} + b_2 I_{dc}^2 \quad (21)$$

where $b_0$= −41.9230, $b_1$=2748.9220, $b_2$= −294.9274. The expression of $I_{dc}$ is defined as $I_{dc}=I_0(1+S_r\cos(\omega_r t))$, where $S_r$ is the distortion factor, $\omega_r$ is the modulation frequency of the anode current ripple. The output voltage function and anode current of a magnetron are always written as:

$$V_{rf} = V_{RF0}\cos(\omega_c t + \theta) \quad (22)$$

Thus, we use the fourth-order Runge-Kutta method to solve these equations by substituting (20), (21) into (10), (18) and (22). The normalized modulation frequency is 0.005. $I_0$ is 0.35 A; $Q_{ext}$ is 400; and $P_{inj}$ is 10 W. The calculated output spectra of free-running and injection-locked magnetrons in the influence of different distortion factors are shown in Fig. 3.

The output phase and voltage of the injection-locked magnetron under various anode current ripples are demonstrated in Fig. 4. We know that the unwanted modulation induced by the ripples of the anode current deteriorates the magnitude and frequency characteristic of the magnetron. The greater the distortion factor is, the stronger the ripple of the magnetron output is. It is mainly caused by the magnetron nonlinearity. Additionally, although injection locking will improve the magnetron's spectrum, the current ripples will still lead to amplitude and frequency disturbance in its output and worsen its phase stability. Therefore, improving the anode current characteristic of a magnetron is important for enhancing the magnetron's performance in either free-running or injection-locked mode.

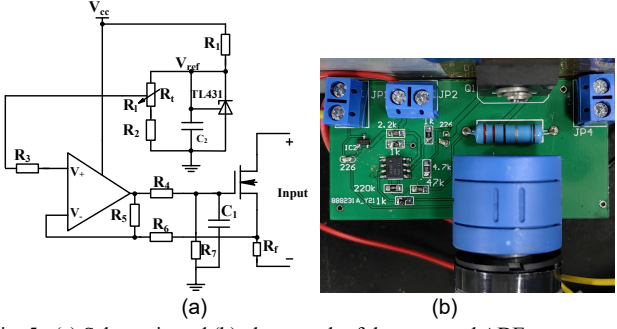

Fig. 5. (a) Schematic and (b) photograph of the proposed ADE.

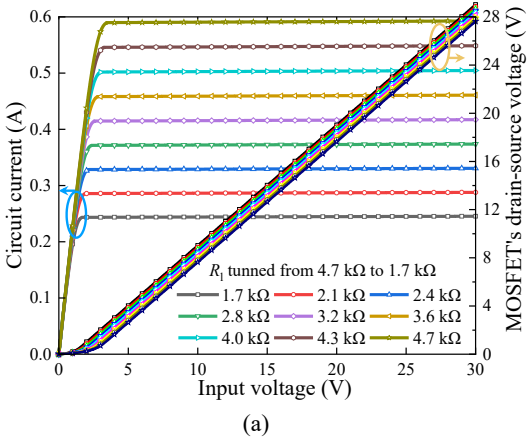

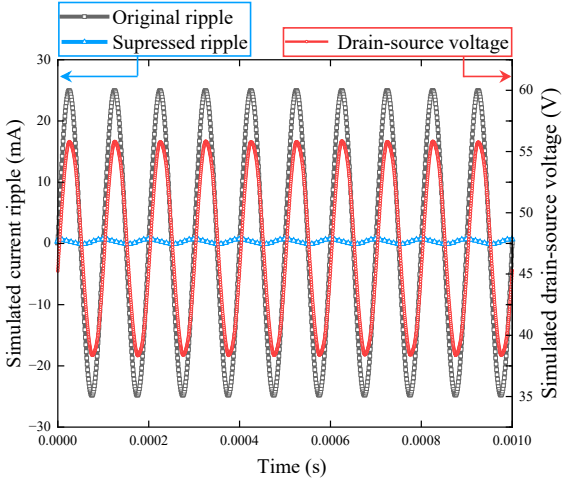

Fig. 6. Simulation results of the (a) output characteristic and (b) ripple suppression effects of the ADE.

### III. REALIZATION OF ANODE CURRENT RIPPLE ELIMINATION

MOSFETs are devices that control current based on the field effects. When the gate-source voltage $V_{gs}$ of an MOS transistor is higher than its threshold voltage $V_t$, and the drain-source voltage is greater than $V_{gs} - V_t$, the MOS transistor enters the saturation region, also known as the constant current region. This MOSFET characteristic can be used to achieve a constant current in active circuits. Therefore, we proposed an ADE built from MOSEFTs.

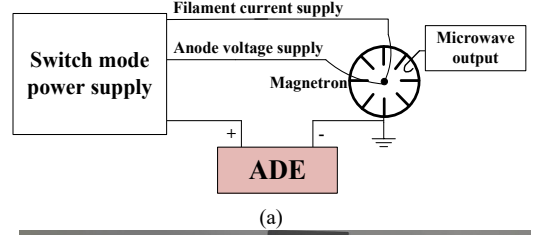

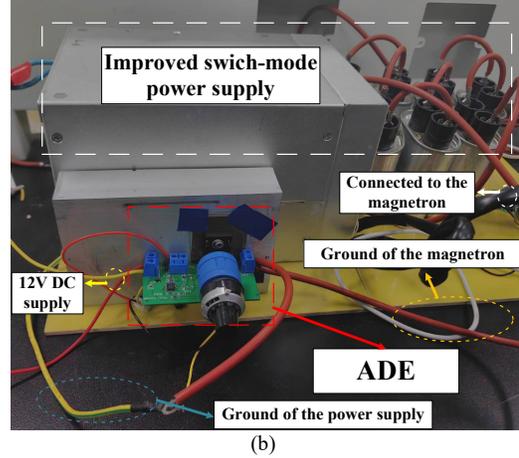

Fig. 7. (a) Diagram and (b) Photograph of the power supply with the proposed ADE.

Its schematic is shown in Fig 5(a). The circuit is designed and simulated in the PSPICE (Orcad Capture). This circuit is a closed-loop feedback circuit. A voltage regulator TL431 provides a reference voltage, and the potentiometer $R_t$ will adjust the input voltage of the operational amplifier's non-inverting input. The sample resistor $R_f$ is used to sample the current and provide a feedback voltage to the inverting input of the operational amplifier. Then, the operational amplifier will give the gate-source voltage to the MOSFET to control the circuit current. Thus, the operation amplifier will compensate for any current fluctuations. An almost constant current control may be realized. If the circuit parameter of the operational amplifier is symmetric, the value of the controlled current is approximated as

$$I_{con} \approx V_{ref} \frac{R_2 + R_l}{R_f (R_2 + R_t)} \quad (23)$$

where $V_{ref}$ is the reference voltage provided by the TL431, $R_l$ is the resistance of the lower section of the potentiometer, and $R_t$ is the total resistance of the potentiometer.

We chose a Panasonic 2M-210 (1000 W, 2.45 GHz) magnetron for experimental verification. Its typical anode current is 360 mA, and the maximum anode current usually does not exceed 420 mA. Thus, we design the ADE parameters to satisfy the anode current requirements. The photograph of the ADE is shown in Fig. 5(b). The operational amplifier is OPA690. A capacitor $C_2$=80 pF is introduced to reduce the harmonic components caused by the OPA690. Besides, in order to ensure linearity of the operational amplifier's output voltage, both $R_2$ and $R_t$ are set at 4.7 kΩ to make the minimum input voltage higher than 1 V. The MOSFET used in the circuit is IRFP460 with a withstand voltage of 500V. The maximum current of the ADE is higher than 420 mA. According to the datasheet of TL431, $C_1$ is set at 22 μF to make $V_{ref}$ 2.5 V. We chose OPA690 as the operational amplifier. Then we preliminary set the remaining circuit parameters based on (23): $R_f$=2.2 Ω, $R_3$=$R_4$=1 kΩ, $R_6$=1 kΩ and $R_5$=220 kΩ. Then, according to the simulation and experiment results, we finally tune $R_6$=47 kΩ to achieve effective feedback control of the gate voltage and a proper current control range.

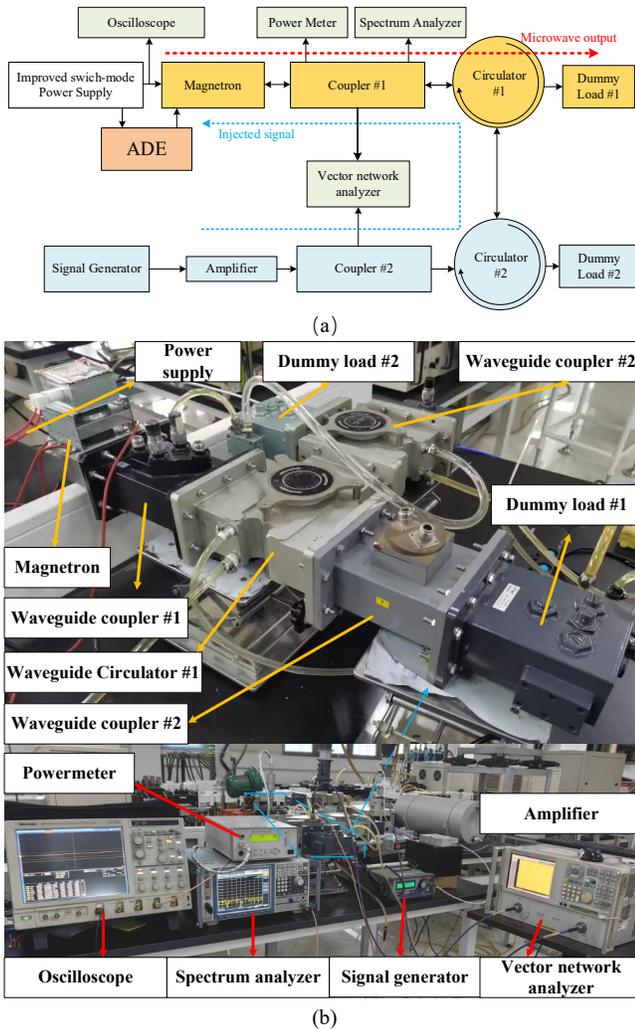

Fig. 8. Setup of the experiment system. (a) Diagram of the experiment system. (b) Photograph of the experiment system.

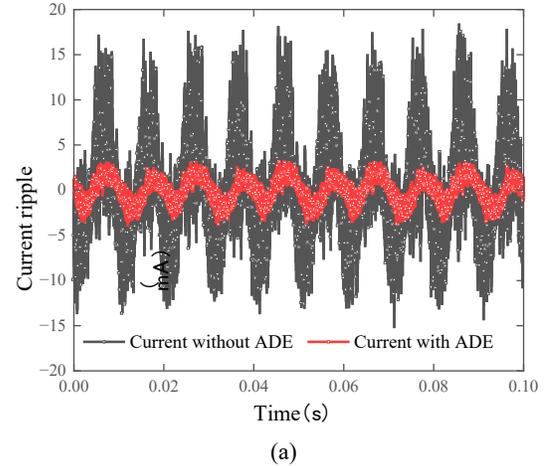

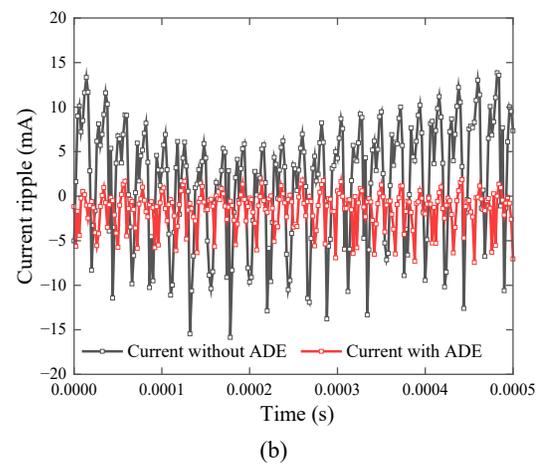

Fig. 9. (a) Low frequency and (b) High frequency current ripples of the magnetron.

In the experiment, initially, $R_1$ will be set equal to $R_t$ to ensure that the MOSFET is completely turned on, corresponding to the maximum controlled current value. This is done to ensure the regular operation of the magnetron. Subsequently, the potentiometer will be adjusted to decrease $R_1$ until the controlled current is comparable to the magnetron's anode current. In this scenario, the MOSFET will control the anode current. As the anode current increases, the MOSFET's gate-source voltage, directly related to the controlled current value, will decrease due to the voltage's negative feedback and vice versa. Consequently, the desired elimination of current distortion is achieved.

Fig. 6(a) presents simulated output characteristics of the ADE. In the simulation, a voltage source with a 2-Ω series resistance provided the input voltage. The simulation results illustrate that the circuit current is controlled by the ADE. When the input voltage increases, the circuit current is restricted at a nearly constant value by the ADE, and the MOSFET's drain-source voltage increases as well. The ADE's current-voltage relationship is similar to the saturation characteristic of the MOSFET. The circuit current is controlled from approximately 600 mA to 240 mA by tuning $R_l$.

Fig. 6(b) gives out the simulated ripple suppression effects and drain-source voltage. The peak-to-peak value of the anode current is reduced from 49.72 mA to 0.86 mA. Moreover, the drain-source voltage $V_{ds}$ of the MOSFET varies simultaneously with the original current ripple, ranging between 38V and 55V. The time-domain waveform of $V_{ds}$ reveals the ADE is operating under the MOSFET's saturation region, aligning with our anticipated behavior. These results indicate that the ADE can eliminate the distortion of the anode current caused by spurious noise and will be applied in the following experiments.

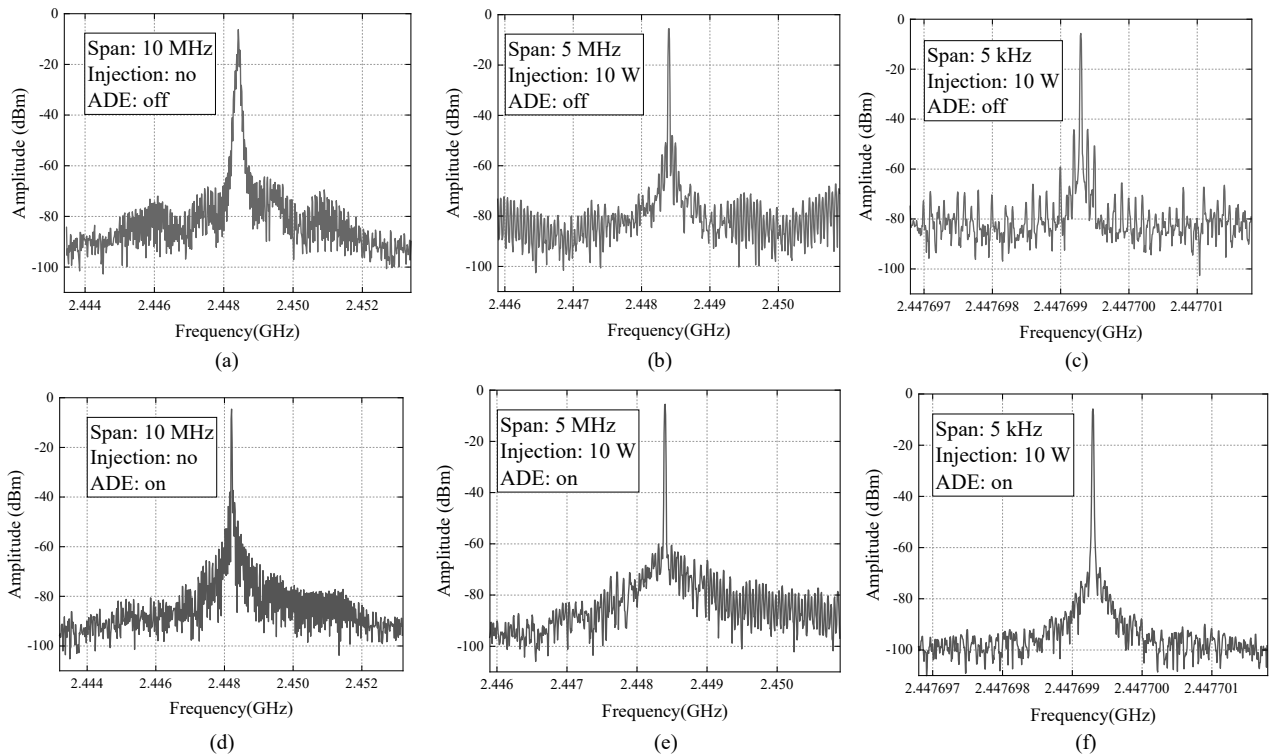

Fig. 10. Spectra of magnetrons in different conditions.

## IV. EXPERIMENT SETUP

An experiment system is established to verify the ADE's effects on the magnetrons' microwave output. The block diagram of the magnetron experimental system used in this paper is shown in Fig. 8(a). Fig. 8(b) shows a photograph of the system. The magnetron is powered by a switch-mode power supply (WELAMP 2000F, Magmeet). A capacitor filter module has been applied to improve the dc output of the switch-mode power supply. The filament current supply will be cut off automatically when the magnetron is working normally. The ADE proposed in this paper is connected between the ground of the power supply and the magnetron, as shown in Fig. 7. The anode current ripple of the power supply is measured using a 1 Ω sampling resistor and an oscilloscope (DPO-7254, Tektronix). The output spectrum of the magnetron is measured by a spectrum analyzer (FSP, R&S), and the output power is measured by a power meter (AV2433).

In the injection locking experiments, a signal generator (HMC-T2220, Hittite) and a high-gain power amplifier (ZHL-30W-262, Mini-Circuits) generate the injection signal. The phase and amplitude stability of the magnetron were measured using a vector network analyzer (Agilent N5230A). We use a water-cooled dummy load #1 to absorb the output microwave power from the magnetron. The reflected microwave will be absorbed by the dummy load #2 to protect the amplifier and signal source.

## V. RESULTS AND DISCUSSIONS

In the experiments, we will first measure the anode current ripple of the magnetron to verify the distortion suppression effects. Then, the output spectrum of the magnetron in its free-running state is measured to correspond to the power supply testing results. Subsequently, the magnetron is injection-locked, and the output spectrum, output phase, and magnitude stability of the injection-locked magnetron with and without the ADE are measured.

### A. Elimination of the Anode Current Ripple

Fig. 9 (a) and (b) show the anode current ripples of the magnetron. The original anode current has a 100-Hz ripple of 31.4 mA, caused by the ac power supply (220V, 50Hz in China) after a full-wave rectifying. The 50-kHz anode current ripple is 21.6 mA, in which the frequency is the switching frequency of the switch-mode power supply. With the ADE enabled, the anode current's low-frequency and high-

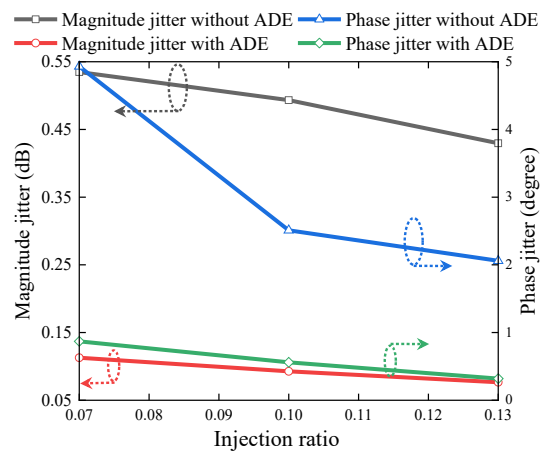

Fig. 11. Measured peak-to-peak value of phase and magnitude jitters of the injection-locked magnetron. (Measurement time: 50ms)

TABLE I
COMPARISON OF PHASE STABILITY OF OUR PROPOSED METHOD WITH OTHER S-BAND POWER SUPPLY RIPPLE REDUCTION METHODS

| Work | Injection ratio | Closed-loop phase compensation | Phase jitter (Time period) | Amplitude jitter (Time period) | Power supply |
|---|---|---|---|---|---|
| [14] | 0.105 | no | ±0.9° (0.1s) | NM | Commercial analog power supply |
| [17] | NM | yes | 0.3° peak-to-peak (0.1ms) ±1° (0.5s) | NM | Home-made switch-mode power supply |
| [18] | 0.13 | yes | Almost ±0.5° (60s) | NM | Improved commercial analog power supply |
| [25] | 0.03 | no | ±9° (NM) | NM | Controllable switch-mode power supply |
| [29] | NM | yes | 0.2° peak-to-peak (0.1ms) and ±0.8° (500ms) | NM | Home-made switch-mode power supply |
| This work | 0.13 | no | 0.32° peak-to-peak (50ms) | 0.07dB (50ms) | Improved commercial switch-mode power supply |

*NM: not mentioned

frequency ripples are reduced to 7.1 mA and 6.7 mA, respectively. These results indicate that the ADE significantly suppresses both the low- and high-frequency anode current ripples caused by the high-voltage dc power supply, which satisfies the design expectation. Besides, based on theoretical analysis, this result also suggests the ADE will significantly improve the performance of the magnetron.

Fig. 10 shows the spectrum of the magnetron under various conditions. We can obtain from Fig. 10(a) that the 30-dB bandwidth is 246.38 kHz for the free-running magnetron without the ADE. The output spectrum presents a multi-frequency and spurious state. When the ADE is turned on, the magnetron output spectrum has been greatly improved. It is similar to the characteristic of a solid-state microwave source, as shown in Fig. 10(d). The 30-dB bandwidth is reduced to 86.96 kHz. Additionally, the measured peak amplitude of the center frequency is increased by 1.74 dB, compared with Fig. 19(a).

The ADE inhibits the spurious components in the magnetron's output, and more energy is redistributed to the center frequency. Experimental results indicate that the ADE can effectively improve the purity and frequency stability of a free-running magnetron's output spectrum. The magnetron's output distortion is eliminated significantly.

### B. Injection-Locking Experiments

Based on the previous analysis, it is known that although the injection signal locks the frequency and phase of an injection-locked magnetron, they are still affected by the current ripples, resulting in spurious frequencies and phase fluctuations.

In the injection-locking experiments, we first set the injection power to 10 W and tune the injection signal's frequency to the free-running frequency of the magnetron. Fig. 10(b) and 10(c) show the magnetron's output spectrum. The magnetron's frequency is locked to the injected signal, and the spurious noise near the carrier is significantly suppressed compared with the free-running state, as shown in Fig. 10(a). However, the spectrum indicates that the magnetron is still affected by the 50-kHz and 100-Hz ripples from the switch-mode power supply. The amplitude of the spurious components caused by the 100 Hz ripples is -49.53 dB/Hz at 100 Hz offset and -55.92 dB/Hz at 200 Hz offset. The 100-Hz related high-order noise components also appear in the spectrum. The noise levels of the 50 kHz and 100 kHz noise caused by the high-frequency ripples of the power supply are -82.88 dB/Hz and -85.80 dB/Hz, respectively.

TABLE II
COMPARISON OF OUR PROPOSED METHOD WITH OTHER METHODS

| Method | Cost | Volume | Complexity |
|---|---|---|---|
| PLL circuit | High | Medium | High |
| Anode voltage filter module | Medium | Large | Low |
| Proposed method | Low | Compact | Low |

When the ADE is turned on, as shown in Fig. 10(e) and (f), the noise of the magnetron is suppressed significantly. The noise levels are -72.81 dB/Hz at 100 Hz, -73.04 dB/Hz at 200 Hz, -98.53 dB/Hz at 50 kHz, and -95.52 dB/Hz at 100 kHz, respectively. The low-frequency noise is almost eliminated, and the high-frequency noise is significantly reduced. All noise suppression strength caused by current ripples is over 10 dB. The highest suppression strength is 23.32 dB.

Because of the non-linearity of the circuit, some harmonic frequency noise appeared. For example, the noise level at 200 kHz offset is increased by 3 dB, as shown in Fig. 10(b) and (e). However, the performance degradation caused by this non-linearity is far less than the improvement achieved. This result indicates that the ADE is effective for magnetron's noise suppression and can be used to partially replace the conventional bulky LC-type filter to eliminate current ripples, especially low-frequency ripples.

Then, we experimented on the magnetron's phase stability. The phase difference was measured using a vector network analyzer. Fig. 11 shows the measured peak-to-peak value of the phase and magnitude jitters of the injection-locked magnetron in different injection power ratios. Both the phase and magnitude jitter decrease with the increase in the injection power ratio.

After the ADE is turned on, when the injection ratios are 0.07, 0.1, and 0.13, the peak-to-peak value of the injection-locked magnetron's phase jitter is reduced from 4.93°, 2.51°, and 2.06° to 0.87°, 0.56°, and 0.32°, respectively. The peak-to-peak value of magnitude jitter is improved from 0.53dB, 0.49dB, and 0.43dB to 0.11dB, 0.09dB, and 0.07 dB. This result indicates that the ADE effectively suppresses the injection-locked magnetron's phase and magnitude jitters. We observe from Fig. 11 that when the ADE is turned on, even in cases with relatively low injection power ratios, better phase and magnitude stability is achieved than in situations with higher power injection ratios without the ADE. The results are consistent with our numerical analysis in Sec. II.

Table I compares the method proposed in this paper and other ripple inhibition methods. It can be seen that the injection-locked system with the proposed technique has phase stability comparable to systems that have PLL circuits [17][18][29]. In this experiment, the magnetron is powered by the improved commercial switch-mode power supply. Magnetron systems with homemade analog power supplies will be established in the future. Better phase and magnitude stability is expected to be realized.

Table II shows the comparison of our proposed method with other conventional methods. The proposed technique has the advantages of being compact, low cost, and low complexity. However, the proposed method also has some limitations. The current ripples of the power supply need to be less than a particular value to avoid the MOSFET's breakdown. Otherwise, the constant current effects affect the regular operation of the magnetron. Besides, the MOSFET's non-linearity response may influence magnetrons' output. In the future, more studies on the proposed method will be conducted to realize better performance and broader applicability.

Experiments verified that the ADE realized output distortion reduction, output frequency stabilization, and phase/magnitude jitter suppression were based on the current characteristic of a magnetron. Besides, the experiment results coincide well with our theoretical estimation. The proposed ADE achieved excellent performance and is expected to partially replace the conventional LC filters used for suppressing anode current ripples.

## VI. Conclusion

This paper proposes a novel method to improve a magnetron's output frequency, phase, and magnitude characteristics based on its anode current characteristics. The impact of the magnetron's anode current on its free-running and injection-locked states is theoretically analyzed and numerically calculated. An ADE designed to improve the output characteristics of the magnetron is developed and experimentally verified.

The ADE suppresses the low- and high-frequency anode current ripples to 22.61% and 31.01% of their original values. The spectrum width of a free-running magnetron is decreased from 246.38 kHz to 86.96 kHz. The maximum suppression strength of an injection-locked magnetron's output noise reached 23.32 dB at 100 Hz offset. Besides, the peak-to-peak value of the phase and magnitude jitter are reduced by 84.47% and 83.72% at the maximum, when the injection power ratio is 0.13, respectively. The experimental results are consistent with the trends in our theoretical analysis and numerical calculations.

This method is expected to be applied in various fields, such as SWIPT, coherent power combining, wireless power transmission, phased arrays, etc., to achieve small-size, low-cost, and high-performance microwave sources. Besides, the proposed method can also be combined with voltage filter modules and PLL circuit methods to realize a more precise phase and magnitude control of a magnetron. The proposed technique provides a novel idea to improve a magnetron's output characteristics with a simple system, compact size, and low cost. It is expected to replace the large volume and high-cost low-frequency capacitors and inductors in ripple elimination.

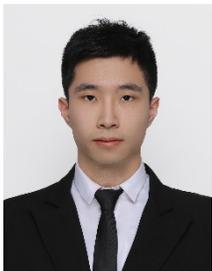

**Shaoyue Wang** (Graduated Student Member, IEEE) received the B.S. degree in Electronic Information Science and Technology from the College of Electronics and Information Engineering, Sichuan University, Chengdu, China, in 2020. He is currently pursuing the Ph.D. degree in radio physics, Sichuan University.

His current research interests include injection-locking technology, high-power microwave sources, and applications of microwave energy.

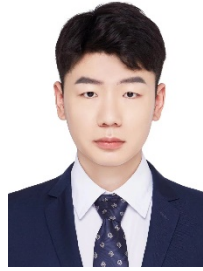

**Yan Zhao** received the B.S. degree in Electronic Information Engineering from the College of Electronics and Information Engineering, Sichuan University, Chengdu, China, in 2021, where he is currently pursuing the M.S. degree in radio physics, Sichuan University.

His current research interests include injection-locking technology and high-power microwave sources.

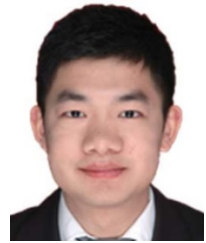

**Xiaojie Chen** received the M.S. degree in radio physics and the Ph.D. degree in communication systems and engineering from Sichuan University, Chengdu, China, in 2018 and 2021, respectively. He has been a Joint-Training doctoral student with the Research Institute for Sustainable Humanosphere, Kyoto University, Uji, Japan, from October 2019 to September 2020. In 2021, he joined the High People's Court of Guangxi, Nanning, China, as a formal staff.

His current research interests include the experimental study of applied microwave engineering and the experimental study of magnetrons.

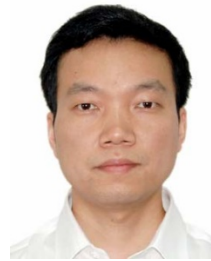

**Changjun Liu** (Senior Member, IEEE) received the B.S. degree in applied physics from Hebei University, Baoding, China, in 1994, and the M.S. degree in radio physics and the Ph.D. degree in biomedical engineering from Sichuan University, China, in 1997 and 2000, respectively. From 2000 to 2001, he was a Post-Doctoral Researcher at Seoul National University, Seoul, South Korea. From 2006 to 2007, he was a Visiting Scholar with Ulm University, Ulm, Germany. Since 1997, he has been with the Department of Radio Electronics, Sichuan University, where he has been a professor since 2004. He has authored one book and more than 100 articles. He holds more than 10 patents. His current research interests include microwave power combining of large-power vacuum components, microwave wireless power transmission (WPT), and microwave power industrial applications.

Dr. Liu was a recipient of several honors such as the outstanding reviewer for the IEEE Transactions On Microwave Theory And Techniques from 2006 to 2010, support from the MOE under the Program for New Century Excellent Talents in the University, China, from 2012 to 2014, the Sichuan Province Outstanding Youth Fund from 2009 to 2012, and named by Sichuan Province as an Expert with Outstanding Contribution. He is the Associate Editor of the Chinese Journal of Applied Science.